\documentstyle[12pt,aasms4]{article}
\begin{document}

\title{STUDY OF VERY SHORT GAMMA-RAY BURSTS}

\author{David B. Cline, Christina Matthey, and Stanislaw Otwinowski}
\affil{University of California, Los Angeles\\
    Department of Physics and Astronomy, Box 951457\\
    Los Angeles, CA 90095-1547 USA\\
    UCLA-APH-0106-4/99 
}

\authoremail{stanislaw.otwinowski@cern.ch}

\begin{abstract}

We have carried out a detailed study of the morphology of gamma-ray bursts (GRBs) with time
duration less than 100
ms that includes: (1) a fast-Fourier spectrum analysis, (2) a comparison with the  Stern analysis
of longer bursts, (3) an inner comparison of the properties of the short bursts, and (4) a comparison
of the short burst properties with the bulk of the GRBs from the Burst and Transient Source
Experiment (BATSE) 4B catalogue. 
We have used the time tagged event (TTE) BATSE 3B data,
which is
available to the public, for part of the analysis. We show that these bursts are very different from the
rest of the GRB events. 
The short bursts appear to be nearly identical, suggesting a separate class of GRBs. We also show
that the short bursts have a Euclidean space--time distribution, in sharp contrast to the longer bursts
with ${\tau_{\rm duration}} >$ 100 ms that implies that these sources are likely local.  Finally we
compare the bursts with a model of primordial black
hole (PBH) evaporation at the quark--gluon (Q--G) phase transition temperature and other shock
wave
models.
\end{abstract}
 
\section{INTRODUCTION}

The physical origin of gamma-ray bursts (GRBs) continues to be unknown. It is clear that the
predominate fraction of the GRBs with $t >$ 1 s have a nearly chaotic luminosity time behaviour
and the events are well described by the Stern time analysis (\cite{Stern96}). It is very likely that
these GRBs are from cosmological sources. For some time, there has been evidence for at least two
classes of GRBs, with a separation time duration of $\sim$ 1 s or perhaps less (\cite{Kouve93}). In
a report (\cite{Cline97}), we reported on our studies of  12 GRBs from the Burst and Transient
Source Experiment (BATSE) 3B catalogue (\cite{Meeg96}) in an attempt to find some regularity.
We have used the BATSE time tagged event (TTE) data in this analysis and performed a fast-Fourier
analysis in these time profiles (\cite{ClineMO98}). Our general conclusion is that the shortest of
these events (with a duration of $<$ 100 ms) behave almost identically, unlike the bulk of the GRBs,
which  possibly indicates
a different physical origin from that of the longer GRBs. We also compare these event distributions
with the bulk of the GRBs to show, by contrast, the great differences in the various sources of the
GRB samples. We briefly explore one such origin, namely, that these GRBs arrive from a
fireball resulting from a primordial black hole (PBH) at the quark--gluon (Q--G) phase transition that
could give rise to such a class of events in Section 8.

The plan of this paper is to first describe the characteristics of the 
short GRBs -- we study the TTE
data for the BATSE 3B catalogue for a detailed fit of the time profile 
and then to discuss various analyses of the events including a fast-Fourier transform analysis and the
resulting power spectrum density (PSD).  Next we discuss some of the problems of getting such
short bursts from current cosmological fireball models.  We then discuss one model that could give
such bursts from PBH evaporation.  Finally, using the BATSE 4B 
catalogue, we compare the ln~$N$--ln~$S$ distributions of the
short GRBs with the bulk of the GRBs where we use a hardness cut to separate the data into different
classes of events.

\section{EVENT SELECTION USING BATSE 3B TTE DATA}

We have studied all events with $T_{90}$ less than 200 ms and then refit the time profile using the
TTE data with the BATSE data set (we restrict this part of the analysis to BATSE 3B data), 
BATSE 3B catalogue (see \cite{Meeg96}). We found 12
events that have a TTE fitted time duration of less than 100 ms. These events are listed
in Table 1 with
the fitted time duration. We then closely inspected the 12 events and found that some have additional
structure. Of the 10 good/fair events, 9 have a time duration of 90 ms or less.
Our goal is to select a similar class of events for study, which constitutes the bulk of the short bursts. 
We only study single peak events, which are the bulk of the short bursts.
As we will show, these events appear to be almost identical in all features, except for BATSE trigger
2463, which we delete in this analysis. This event has a clearly different energy spectrum from the
bulk of the short GRB events. This sample constitutes the set of events studied here.
We show the time profiles in different photon energy bins in Figure 1.

\section{ FOURIER TRANSFORM AND THE POWER SPECTRUM DENSITY}

We have carried out a fast-Fourier transform of all events with $T_{90}$ duration of less than 200
ms. We report here the study of the PSD for the eight events with time duration of less than 66 ms
(see
Figure 2) discussed above. Note the remarkable similarity of the PSDs for these events.

   In the general study of all GRBs, there has never been any evidence of two bursts being alike.  In
fact, the nature of the GRB source is likely to give a chaotic behaviour. However for the eight events
in the sample reported here, there is clear evidence for a high frequency component in the PSD
spectrum.


\section{COMPARISON WITH THE STERN ANALYSIS}

While there are no exact models for any GRBs, there have been some successful phenomenological
models that fit the time distribution on average. One such analysis was carried out by B.E. Stern
(1996).
In this model the GRBs are assumed to fit an analytical model after the first peak of the GRB:
\begin{eqnarray}
I \sim e^{-(t/t_{0})^{1/3}}\,\,\, ,
\end{eqnarray}
where $t$ is the time after the peak and  $t_{0}$ is a constant. Stern showed that the GRBs with
duration greater than about 250 ms fit this distribution very well. In a model-dependent way, one can
associate this behaviour as a sort of cooling-off phase after the initial burst, which seems also to be
the case for solar flares (although with a different $t_{0}$ constant). As far as we know, this is the
most successful analytical description of the general GRB time profiles.
 The results of our analysis of the sum\footnote{In the sum, events are shifted to put the maxima at the same time.} of all 12 short bursts 
(see Table 1) using the analytical form (1) are shown in Figure 3.  
The $t_{0}$ 
value obtained from the fit (0.0016 $\pm$ 0.0007 s) is far from the 
value 
(between 0.3 s and 1.0 s) found by Stern and the fit 
($\chi^{2}$/ndf=770/219)
is not good.  
The failure of the Stern function fit ($\chi^{2}$ of 770 for 219 
degrees of freedom) indicates that the short GRBs have a quantitatively 
different shape from the long GRBs.  In essence, the short GRBs are 
much more symmetric in the rise and fall of the pulse than are the long 
GRBs.
We believe that this is an indication that these very
short bursts are a different class of phenomena from the bulk of the 
GRBs.

\begin{deluxetable}{ccccl}
\tablecaption{{\sc Event Selection and Properties (BATSE 3B) \label{tbl-1}}}
\tablewidth{0pt}
\tablehead{
\colhead{Trigger} & \colhead{Duration from} & \colhead{Hardness} & \colhead{   }&
\colhead{Used in This}\\
\colhead{Number}   &     \colhead{TTE Fit (s)}             & \colhead{Ratio} & \colhead{ 
~~~~~Comments~~~~
} & \colhead{Analysis}}

\startdata
01453       & 0.006 $\pm$ 0.0002      & 6.68 $\pm$ 0.33      & Poor event, precursor & ~~~~No \\
\hline
00512      & 0.014 $\pm$ 0.0006       & 6.07 $\pm$ 1.34      & Good event & ~~~~Yes \\ \hline
01649 & 0.020 $\pm$ 0.0080 &                 & Fair event* & ~~~~Yes \\ \hline
00207 & 0.030 $\pm$ 0.0019 & 6.88 $\pm$ 1.93 & Good event  \\ \hline
02615 & 0.034 $\pm$ 0.0032 & 5.42 $\pm$ 1.15 & Good event & ~~~~Yes \\ \hline
03173 & 0.041 $\pm$ 0.0020 & 5.35 $\pm$ 0.27 & Poor event, precursor & ~~~~No \\ \hline
02463 & 0.049 $\pm$ 0.0045 & 1.60 $\pm$ 1.55 & Good event & ~~~~No \\ \hline
00432 & 0.050 $\pm$ 0.0018 & 7.46 $\pm$ 1.17 & Good event & ~~~~Yes \\ \hline
00480 & 0.062 $\pm$ 0.0020 & 7.14 $\pm$ 0.96 & Good event & ~~~~Yes \\ \hline
03037 & 0.066 $\pm$ 0.0072 & 4.81 $\pm$ 0.98 & Good event & ~~~~Yes \\ \hline
02132 & 0.090 $\pm$ 0.0081 & 3.64 $\pm$ 0.66 & Good event & ~~~~Yes \\ \hline
00799 & 0.097 $\pm$ 0.0101 & 2.47 $\pm$ 0.39 & Fair event* & ~~~~No \\ \hline \hline
      &  & &       *Fair event, small additive structure & \\ 
\enddata

\end{deluxetable}

\section{ COMPARISON OF THE PROPERTIES OF THE EVENTS}

As far as we can tell, no two long duration bursts are alike. We find that this is not true for the very
short bursts, which seem to be identical except for a factor of two in time duration. We
illustrate this in Figure 1, which shows the time distribution, without 
the lowest ($<$ 50 keV) energy,   for the eight very short bursts
selected from the BATSE 3B TTE data, 
and in Figure 4, which shows the energy distribution of 
eight short bursts. 
We include this plot that shows that the energy distributions as well as the hardness are nearly
identical for these events.
This fact can also be
learned from the hardness values in Table~1 (Cline et al. 1997).
Within error, all have the same
hardness (excluding BATSE trigger 2463). To our knowledge there is no evidence for any other set
of GRBs with such identical features. We believe that this indicates a simple and identical
production process for this class of events.

\section{DETECTION EFFICIENCY FOR SHORT BURSTS}

The bulk of the very short bursts identified here all have time duration at or below the BATSE 64-ms
integration
time. We therefore believe that the BATSE trigger is likely an inefficient method of  identifying such
events; we also believe that many weak  bursts may have been  missed.

Recently the issue of detecting  short GRBs with time duration of $\tau <$ 64  ms (the smallest
BATSE trigger time scale) has been raised (\cite{Nem98}).
They show that the detector
efficiency will drop sharply for bursts below 64 ms. Since several GRBs have been detected  with
bursts less than 64 ms, it is likely that there is a significant population of short bursts that have been
missed. Here we will
not attempt to determine the detector efficiency as a function of $\tau$ but simply refer to the work
of Nemiroff et al. (1998).  They state that there could be as many missed GRBs of 1-ms
duration as the number that have been detected at 10 s. 

Since the detection of the short bursts is uncertain, we may not expect as clear a separation of this
class
of events from the bulk of the GRBs as from the two classes of GRBs that have 
been identified (\cite{Kouve93}).

\section{BEHAVIOUR OF INDIVIDUAL EVENTS}

\subsection{BATSE Trigger Number 512}
To obtain a better understanding of the short bursts, we discuss three individual bursts that have
more
detailed information than the bulk of the events. According to the arguments given above, we expect
all of the short bursts to be very similar and, therefore, we assume that the behaviour of these special
bursts is likely an example for all short bursts. If these events are typical of the short bursts, then we
can see a clear behaviour in the fine time structure and the detected gamma energy distribution.
We start with the incredible GRB trigger 512. In Figures 5 and 6, we note the detailed fine structure
for BATSE trigger 512, which has the finest time structure of any GRB observed to date --
possibly down to 20 $\mu$s level.  This was a very bright burst and allowed unpredicted time
information.

\subsection{PHEBUS Event Numbers PB900320 and PB900813}   
The PHEBUS GRB detector [see PHEBUS catalogue ({\cite{Terek94})], has recorded two very
interesting short time events, shown in Figure 7. As far as can be determined, these events are
identical (note that the energy distribution are fit to a synchrotron and are identical). 
Because of a thicker absorber, the PHEBUS detector has a larger energy-range capacity than that of
the
BATSE detector.

Thus this detector can record photon energies up to 180 MeV in contrast to BATSE, which is only
sensitive up to about 580 keV due to the the absorbers. Note that these two PHEBUS events have
photon energies above 1 MeV. Thus the short GRBs have energetic photons in the spectrum.

\section{EXPECTATIONS FOR SUBSTRUCTURE IN FIREBALL MODELS}

   As we have shown, the very short bursts analyzed here have strong substructure on the scale of 50
$\mu$s in at least one case and pulse rise times of the order of milliseconds. We now explore the
models for GRBs of cosmological origin.  For the fireball models with total energy release
of 10$^{51}$ ergs, it is very difficult to get such short bursts, because it requires very
large Lorentz boosts for the shock front. In these same models, the possibility
of temporal substructure has been studied recently by Panaitescu and M\'{e}sz\'{a}ros (1998) and
they found that it describes the important characteristics of the bulk of the GRBs in the BATSE
sample rather well.

   In one fireball model, the burst time scale is set by the hydrodynamic time scale $t_{\rm dec}$ and
the fireball Lorentz factor is $\Gamma_{0}$ = $E_{0}$/$Mc^{2}$, where $E_{0}$ is the total energy
released and $M$ is  the entire baryonic mass in the fireball, which gives $T$ = 10 $\cdot$ $t_{\rm
dec}$/2 $\cdot$ $\Gamma^{2}_{0}$. The BATSE data can be explained by $\Gamma_{0}$ values of
$\sim$
100--500, and $E_{0} \sim 10^{50}$ erg gives $T \sim$ 2--20 s to  get very short bursts requiring
a very large value of $\Gamma_{0}$. For temporal substructure to occur, it must be related to even larger
values of $\Gamma_{0}$ and other factors that we will not discuss here. In the case of BATSE  trigger 512
with $\Delta t \sim$ 50 $\mu$s, the Lorentz factor must be extremely high (50,000), which seems
to be very unphysical in this model. This model also predicts very asymmetric rise and fall times for
the bursts that are very similar to those of the Stern analysis discussed before. As we have shown,
the very
short bursts are rather symmetric. We conclude that it is unlikely that the very shorts bursts can be
described by a cosmological fireball model.   

   We also consider the effects of internal shocks, as has been discussed recently by Kobayashi,
Piran, and Sari (1997) and references included therein.  
In this model, the source releases
energy of $10^{52}$ erg with negligible baryonic contamination $\leq 10^{-5} M_{\odot}$.  The
region of GRB emission is found to be $\sim 10^{11}$ cm and fireball shells collide with
distances of $ \sim 10^{14}$ cm, generating the internal shocks and the time variation in the GRB. 
The natural times for such GRBs are
$r/\Gamma^{2}_{0}c
\sim 10^{14}/(10^{2}\times3\times10^{10}) \sim 30$ s,
which are far beyond the time scale of the short bursts studied here.  The time scale of the internal
shocks is 
$\sim 10^{11}/\Gamma^{2}_{0}c \geq 3$ ms,
which is far longer than the 50 $\mu$s structure observed for BATSE trigger 512.  We conclude that
cosmological fireball models with internal shocks are unlikely to produce events of the type studied
here.  The main reason is that the enormous energy released requires a large volume for photon
emissions, which implies longer time scales for the overall burst and variation. We therefore
conclude that it is very unlikely that these GRBs are due to cosmological sources.

\section{ONE POSSIBLE MODEL FOR  THE ORIGIN OF THE VERY SHORT  BURSTS}

Ever since the discussion of PBHs began, there have been suggestions for the experimental detection
of such objects. However, the real-time detection depends on the final state evolution of the PHB;
as it sheds mass, the temperature of the PHB rises into the region of hadronic interactions and
hadronic final states. The most extreme model used to simulate this final  state was the Hagedorn
model, which predicted an explosion lasting $\sim$ 10$^{-7}$ s, whereas other QCD-inspired
calculations suggested a final-state collapse time of the order of seconds. We questioned the validity
of the QCD-inspired calculations, pointing out that final state interactions and non-perturbation
effects could increase the low-energy particle luminosity and even the collapse time, both of which
would make detection easier. We note that there are already indications of two classes of
GRBs (\cite{Kouve93}). The approach taken here is to determinate if there are GRBs that could be
consistent with PBH evaporation, not to attempt to prove the existence of PBHs at this stage.
Hawking (1974) showed that the temperature of the PHB increases as it loses mass during its
lifetime. Mass, in the context of the standard model of particle physics,  is lost at a rate
\begin{eqnarray}
\frac{dM}{dt} = -\frac{\alpha(M)}{M^{2}}\,\,\, ,
\end{eqnarray}
where $\alpha(M)$, the running constant, counts the particle degree of freedom in the PBH
evaporation. The value of the $\alpha(M)$ is model dependent (Cline et al. 1997; Hawking 1974).

   Black holes at the evaporation state in the present epoch can be calculated as
\begin{eqnarray}
M_{*} \simeq [3\alpha(M_{*})\tau_{\rm evap}]^{1/3} 
\simeq 7.0 \times 10^{14} {\rm g} \,\,\, , 
\end{eqnarray}
where $\alpha(M_{*}) \simeq 1.4\times10^{-3}$. Thus, the number of black holes with critical mass
$M_{*}$ in their final state of evaporation is
\begin{eqnarray}
\frac{dn}{dt} = \frac{3\alpha(M_{*})}
{M_{*}^{3}}N = 2.2\times10^{-10}N 
{\rm pc}^{-3}{\rm y}^{-1}  \,\,\,  ,
\end{eqnarray}
where $N$ is the number of PBHs per pc$^{-3}$ in the Galaxy.

   In previous works (Cline \& Hong 1992; Cline et al. 1997)
it was shown that short GRBs have the characteristics
expected for PBH evaporation.
In these same references, it has been shown that the Hawking 
Radiation can be detected as a short GRB from PBHs as far away as a 
few parsec, because of the enhancement from the quark--gluon phase 
transition, provided it is a first order transition.  In this case, 
and using constraints on the PBH density universe in these references, 
we would expect a few events per year to be detected by BATSE.  
This rate is consistent with the rate of short GRBs studied in this 
paper.

\section{COMPARISON WITH OTHER GRBs}

We classify all GRBs into three different categories: 
one with $\tau > 1$ s (long, L), one with 1 s $> \tau >$ 0.1 s (medium, M), 
and  one with $\tau \leq$ 100 ms (short, S), 
which is the focus of this investigation. The location of
these events is shown in Figure 8.  
In Figure 9 we show the contrast between the S and M classes of events, which appears
to be counter intuitive.

We note that the short bursts are strongly consistent with a ${{C_{p}}^{-3/2}}$ spectrum,
indicating a Euclidean source distribution, as was shown previously by (D. Cline et al. 1997).
In the medium (100 ms to 1 s) time duration, the ln $N$--ln $S$ distribution seems to be non-Euclidean;
in the long duration ($\tau >$ 1 s) bursts, the situation is more complicated as we have shown
recently (Cline, Matthey, \& Otwinowski 1998b).  The $<V/V{\rm_{max}}>$ for the S, M, and L
class of events is, respectively, 0.52 $\pm$ 0.1, 0.36 $\pm$ 0.02, 
0.31 $\pm$ 0.01.  (In this case, we have used the BATSE 5B 
data\footnote{We understand BATSE 5B data to include all BATSE data
registered before Nov. 18, 1998.}
to obtain the best statistics.)

   We point out again that the 12 events discussed in Sections 2, 3, 
and 4 are obtained from the short GRBs in the BATSE 3B catalogue, 
as explained in Section 2; the larger sample of short GRBs used in 
Section 10 came from the most recent data up to Nov. 18, 1998 
(beyond the BATSE 4B catalogue).  In the latter case, we needed to 
collect the largest statistics for the results in Figure 9. 

  One possibility for explaining these effects is that the distribution of the short bursts
may come from a local Galactic source.  This explanation is not viable for the medium time
bursts, since $<V/V{\rm_{max}}>$ is 0.36 $\pm$ 0.02, indicating a likely cosmological source. 
The longer bursts are clearly from cosmological distance.

By studying Figures 1--3, it can be seen that these events could well be nearly identical except for
different time durations, and the identical energy spectra for the two PHEBUS events (Figure 7)
gives 
additional support. In addition, the ln $N$--ln $S$ behaviour (Figures 8 and 9) suggests
an extremely compact source. These are all consistent with PBH evaporation as the source of  GRBs
and imply a Galactic origin of the short GRBs. Of course there could be other types of Galactic
sources as well. 

In summary, we have shown that the GRBs with $\tau$ $<$ 100 ms likely are due to a separate class
of sources and appear to be nearly identical in contrast to the bulk of GRBs.  
     In this analysis, we have studied a small class of BATSE 3B TTE data in detail, and to improve
the statistical power for some issues, we have used BATSE 4B and the latest BATSE 5B data.  We do not believe
this study warrants the use of the full TTE data for BATSE 4B or 
BATSE 5B, since the point is to show a general
morphology of the GRBs, not a complete statistical analysis at this stage.
It is likely that the source is local or Galactic, in contrast to the cosmological origin of the bulk of
GRBs.  One model source that may produce such a unique class of GRBs is the evaporation of
PBHs.  Independent of that model, we believe these short bursts constitute a third class of GRBs.

\section{Acknowledgment}
We wish to thank members of the BATSE team for discussions, especially J. Norris and G. Fishman.



\clearpage
\noindent Figure Captions:

\figcaption{ Time profile for the eight short GRBs in different energy bins from the BATSE 3B TTE
data.\label{Figure 1.}}
\figcaption{ The power spectrum density of eight short GRBs using BATSE 3B TTE
data.\label{Figure 2.}}
\figcaption{The result of applying the statistical Stern analysis to the sum of 12 short GRBs.\label{Figure 3.}}

\figcaption{ The energy spectrum of the eight events discussed in the text plotted as a function of
time.\label{Figure 4.}}


\figcaption{ Plot of the rise time structure for GRB trigger number 512.\label{Figure 5.}}

\figcaption{ Plot of the fine time structure for GRB trigger number 512.\label{Figure 6.}}

\figcaption{Two short bursts from the PHEBUS detector:  
Time profiles of (a)PB900813 and
 (b) PB900320, and energy spectrum of (c) PB900813 and   
(d) PB900320.\label{Figure 7.}}

\figcaption{Hardness (H) vs duration (S = short, M = medium) of 
BATSE-4B
GRBs.\label{Figure 8.}}

\figcaption{ Comparison of the ln$N$--ln $S$ distribution of the short (S) and medium (M) 
GRBs for BATSE 4B data.\label{Figure 9.}}

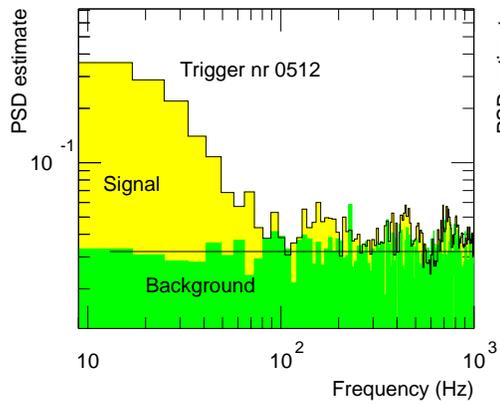
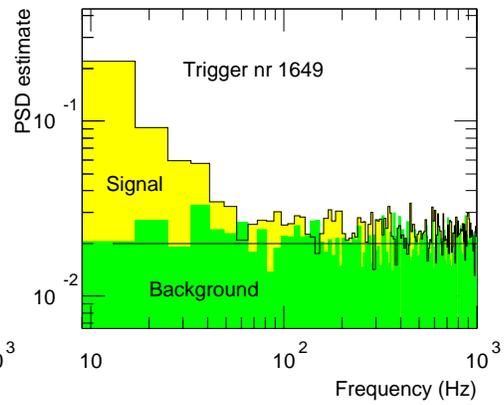
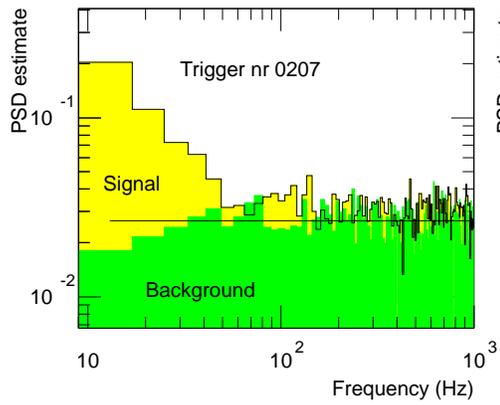
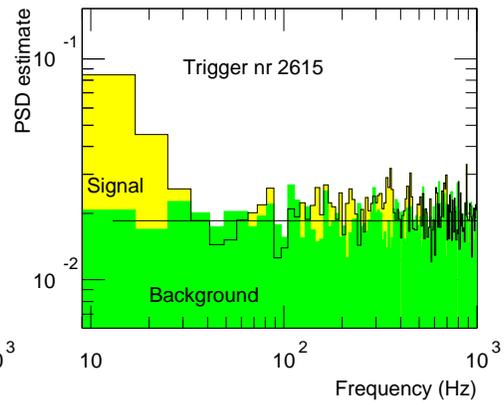
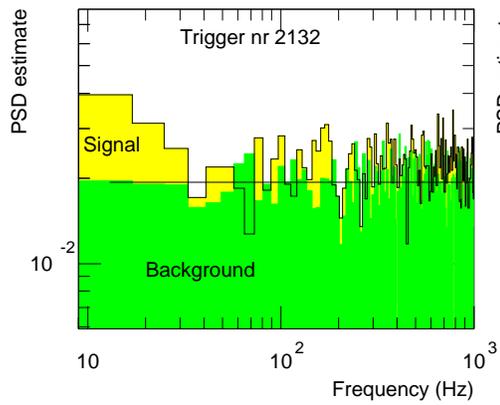
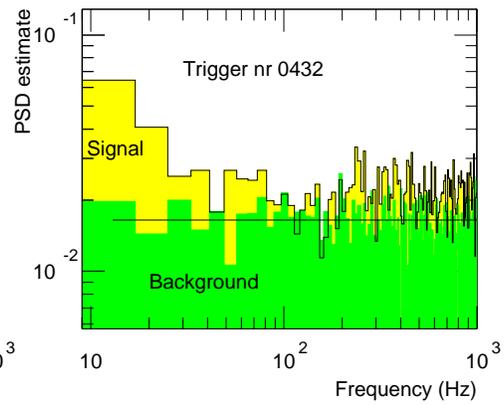
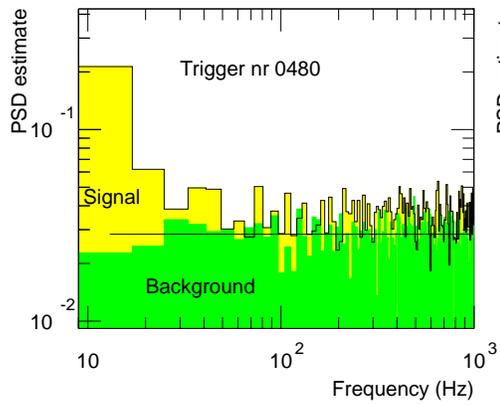
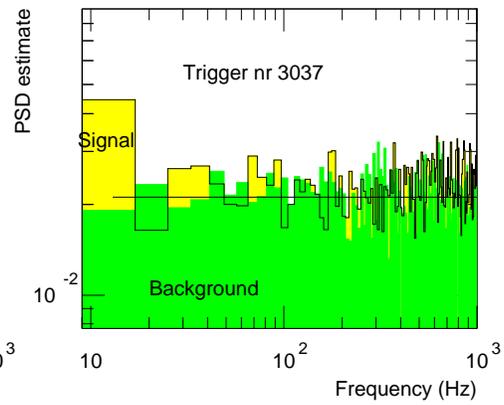
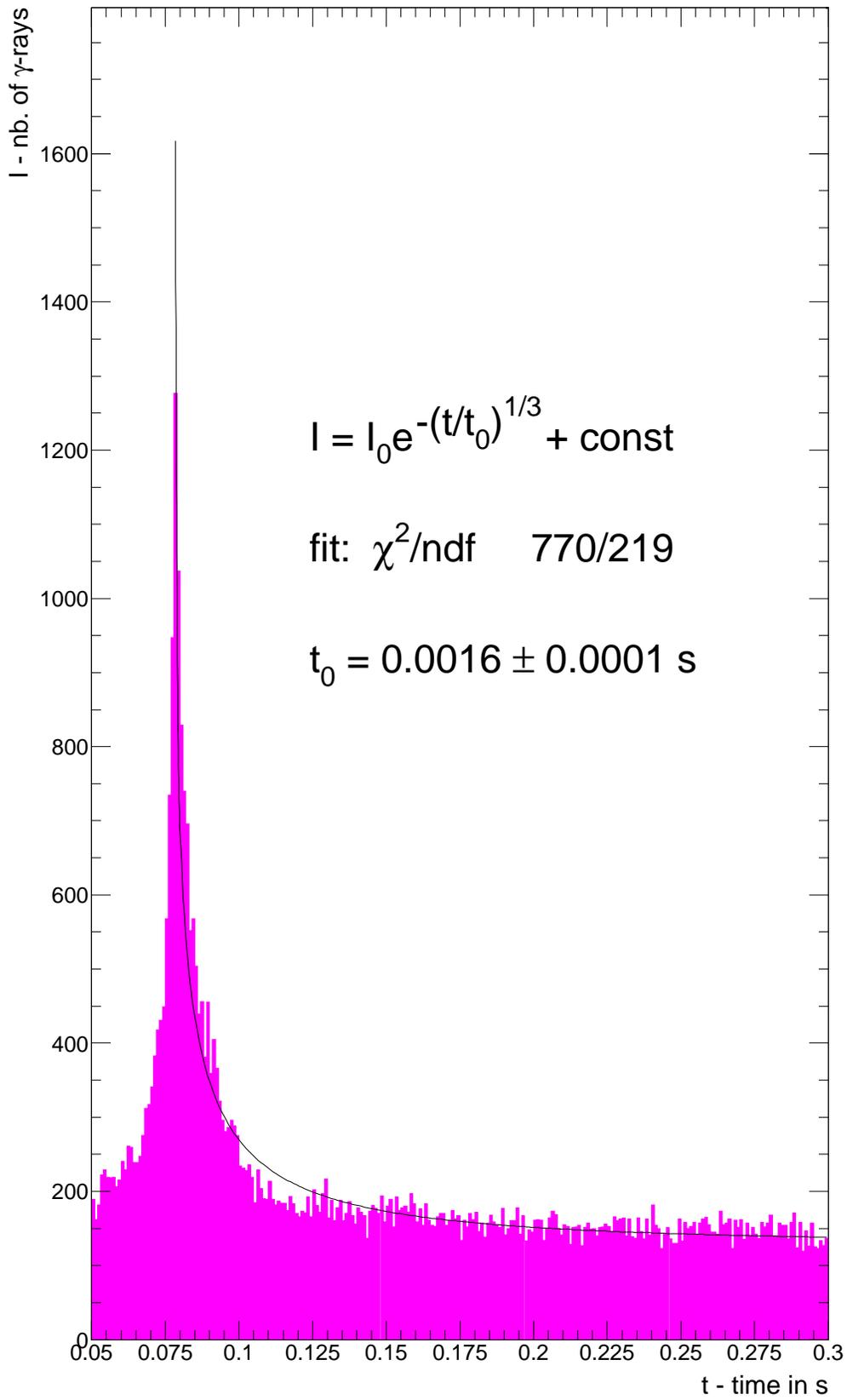
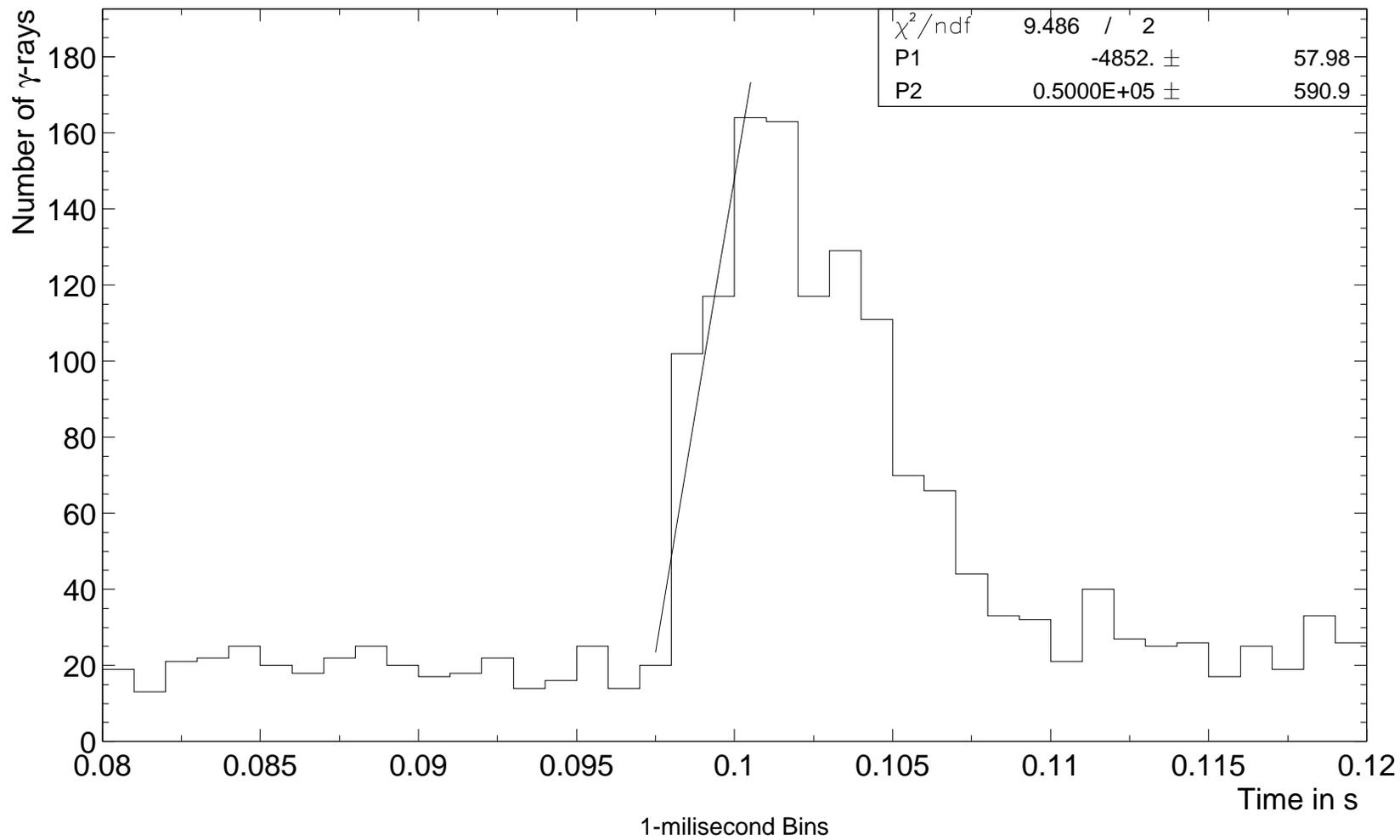
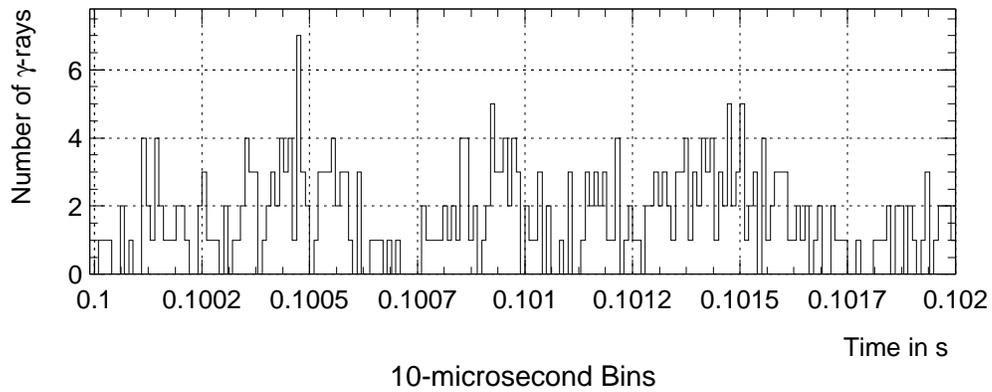
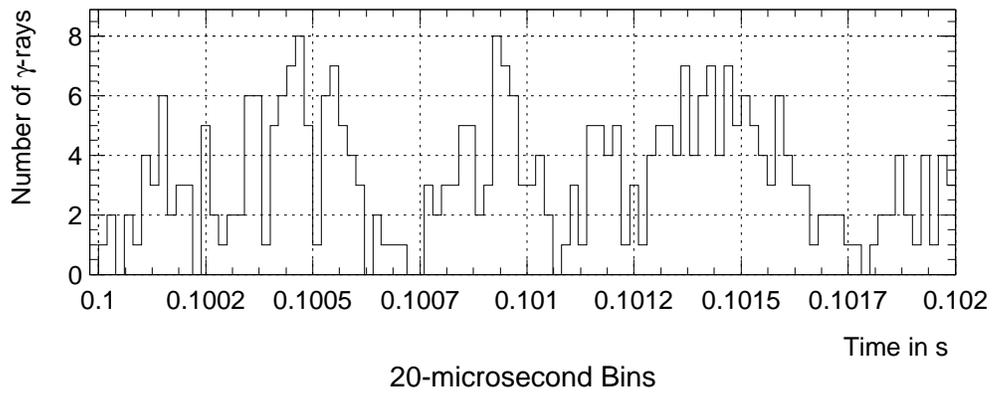
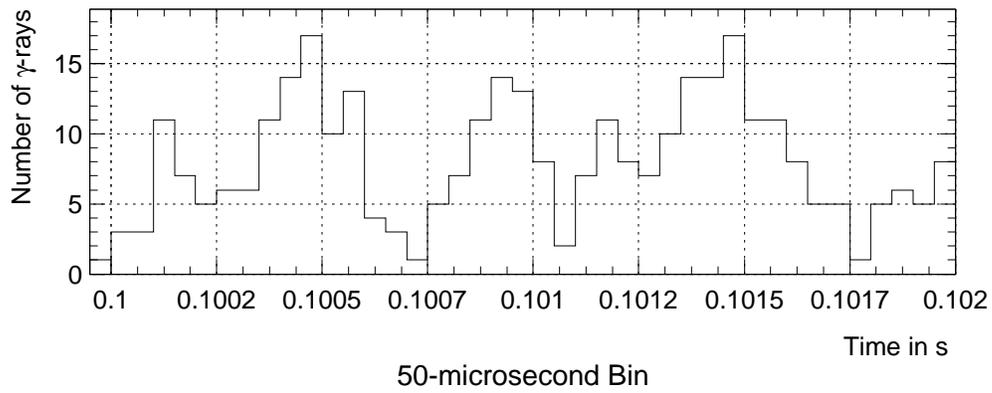
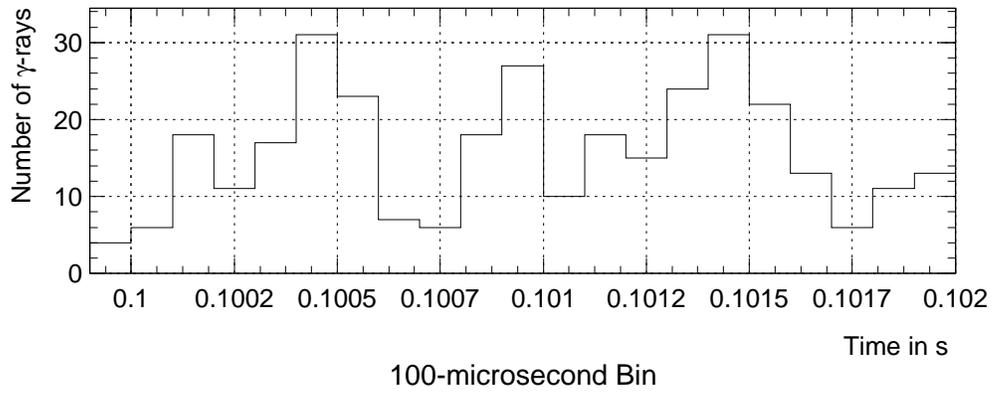
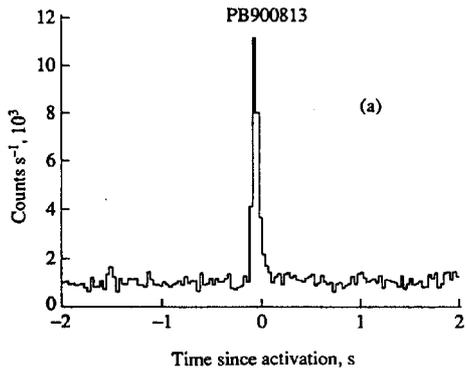
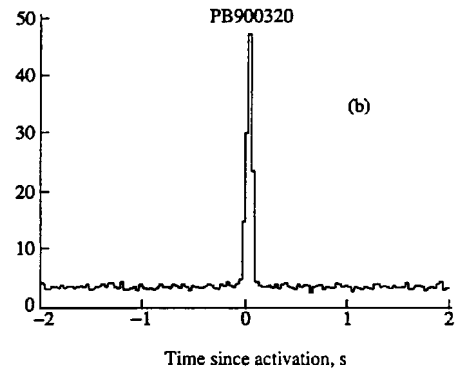
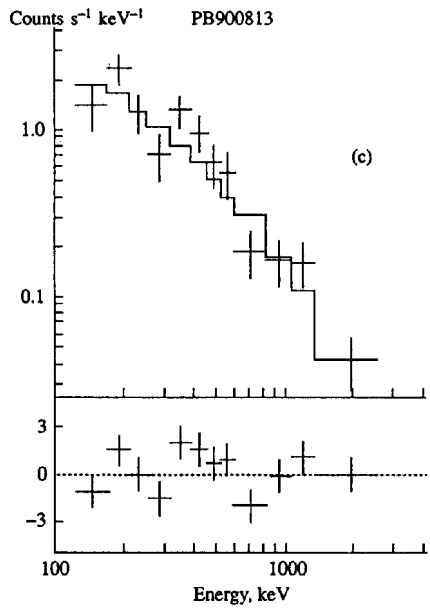
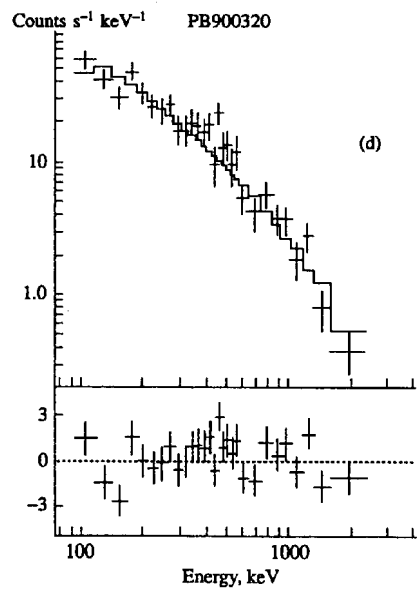
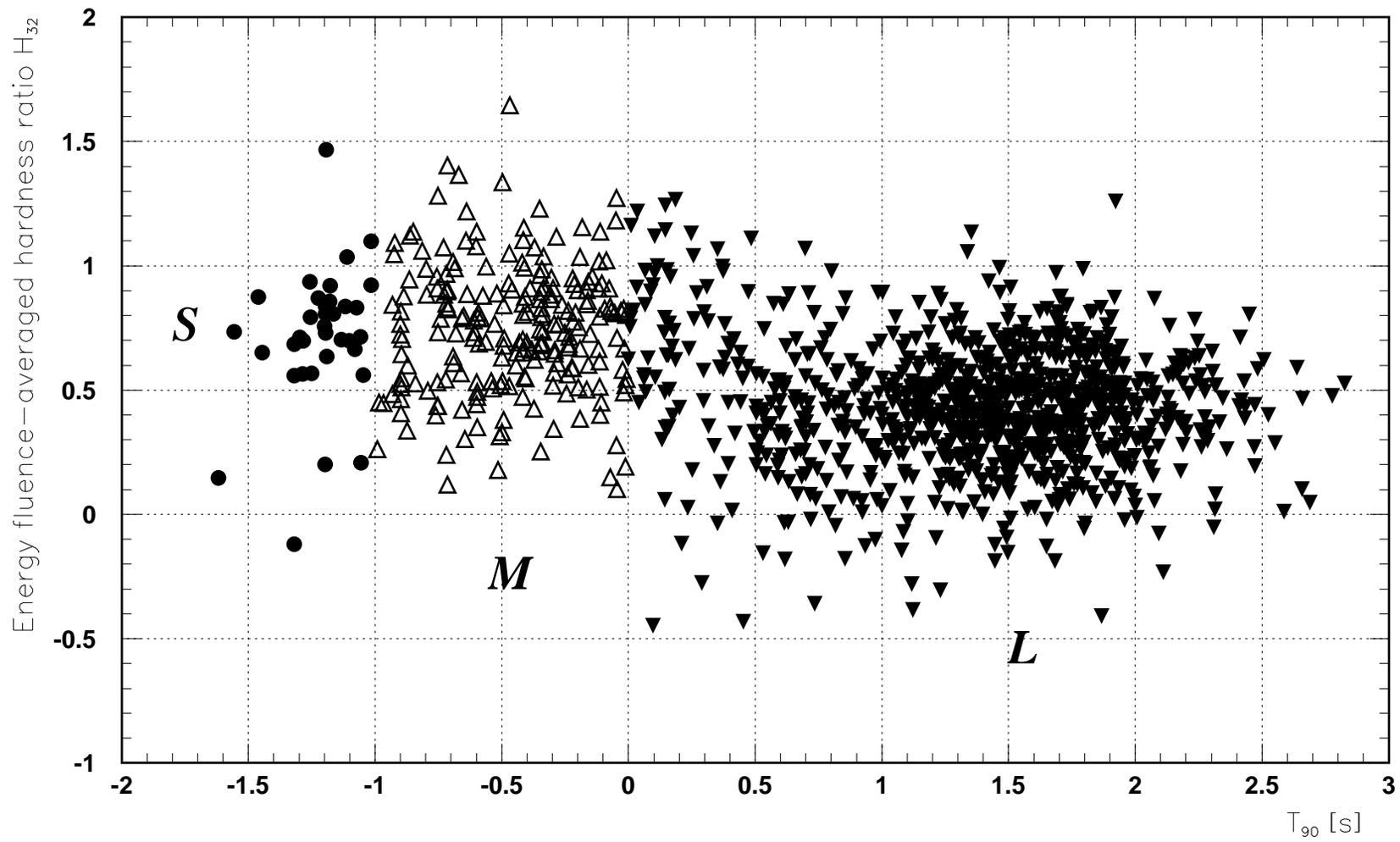
\end{document}